\documentclass[twocolumn,secnumarabic,amssymb,showpacs,nobibnotes, aps, prd]{revtex4-1}

\usepackage{epsfig}
\usepackage{graphicx}
\usepackage{amsmath}
\begin{document}

\title{Raman and non Raman electromagnetically induced transparency  resonances  in a degenerate four level system}

\author{Haji Ahmedov }
\email{haji.ahmadov@tubitak.gov.tr}
\author{ Cengiz Birlikseven }
\affiliation{TUBITAK, UME (National Metrology Institute),
Gebze Yerleskesi, 41470 Gebze, Turkey}

\begin{abstract}
We study  coherence effects in an atom  having three metastable levels and single  life-time broadened upper state. Two laser fields couple ground states to an excited one and an radio-frequency (rf) field operates between ground state levels. We consider the case when the common level for the rf-field and one of the laser field ( to which we refer as  the probing field ) transitions is twice degenerate.  We predict a novel type of dark states ( we call them non Raman dark states ), which in contrary to Raman resonances, do not require the fulfillment of the two-photon  resonance condition.  We realize the configuration  in the atomic hydrogen and show that  non Raman resonances  are determined by the geometry of the rf and probing  field's polarizations. For ultra-cold atoms non Raman resonances arise when the polarization vectors of the fields are parallel or perpendicular to each other. For  an atomic gas at room temperatures  electromagnetically induced transparency ( EIT ) resonances associated with  Raman and non Raman dark states are studied by taking into account  spin-exchange  relaxation  in the ground state of the hydrogen atom.   We establish properties of non  Raman dark states   by investigating  the relation  of   EIT resonances to the polarization geometry.

\end{abstract}
\pacs{ 42.50.Gy,42.50.Ar,42.50.Ct}
\maketitle

\section{Introduction}

Coherent population trapping (CPT) is one of the remarkable phenomenon associated with  quantum  coherence and interference in a three-level atom. The essence of coherent population trapping is that, under certain conditions, there exists a coherent superposition among atomic state ( the dark state ), which is decoupled from the coherent and dissipative interactions \cite{Arimondo}. When an atomic system is prepared in the dark  state  electromagnetically induced transparency (EIT) can be observed where the atomic coherence cancels or reduces absorptions  \cite{EIT, EIT1}.  The importance of EIT resonance  stems from the fact that it gives rise to greatly enhanced susceptibility (  the real part of the susceptibility ) in the spectral region of induced transparency of the medium  which leads to the enhancement of the refractive index   \cite{Slowlight1,Slowlight2}. Another interesting nonlinear effect related to dark states is the adiabatic population transfer, which allows to  prepare the atom in a desired state and to control transitions between atomic levels by means of adiabatically changing continuous radiation waves \cite{ADIABATIC}.  The interest to dark states is due not only  to the fascinating physics involving quantum interference but also to the fact that there are many potential applications such as lasing without inversion \cite{LWI0, LWI1,LWI2},  quantum information processing and storage  \cite{QINF}, subrecoil laser cooling \cite{lasercooling} and atom interferometry \cite{interf}.

\noindent 

The  dark  state was observed for the first time by  G. Alzetta et al. \cite{CPT1} in the Lambda ( $\Lambda$ ) configuration  in  which   two ground  states are coupled to a single  excited  state by a bichromatic field. The single dark  state appears when two laser fields  satisfy the two-photon Raman resonance condition for the transition between  two  ground states. It is possible to enlarge the domain of the  dark  state based physics by adding an additional metastable level to the Lambda system. In this case  double dark state structure is observed in the configuration in which the additional level is coupled by a rf field to one of the Lambda system's ground state \cite{RfExp,DDS2,DDS3,DDS4}. The appearence of double dark state structure is associated with  the Autler-Townes splitting of atomic levels induced by the rf field.  The resonances associated with the double dark states can be made absorptive or transparent and their optical properties such as width and position can be manipulated by adjusting the coherent interaction  \cite{DDS1}. In the configuration where three laser field operates between ground states and excited one,   two coupling laser fields  can be used  for slowing down one weak probing field. As a result  two dark states appear in the system which lead to two EIT resonances and two group velocities for the probe field  \cite{Tripod}. Along with the Lambda system it is also of great interest atomic  coherent states and associated nonlinear optical effects in a so-called Vee ( $V$ ) and Cascade ( $\Xi$ ) configurations  where more than one level may be unstable. In recent years, many extensive studies have been devoted to  four-level systems which are the generalization of these three-level configurations \cite{F1,F2,F3,F4,F5}.

\noindent 

This work is adressed to the study of coherence effects in the four level system which one metastable level is twice degenerate ( the configuration is shown in Fig. 1(a) ).   The rf field induced dynamical Stark effect splits  the degenerate level into three sub-levels. As a result three dark states appear when  the coupling and the probing fields satisfy the two-photon  resonance conditions  between corresponding lower state's transitions ( see Fig. 1(b)). This triple dark state structure is the straightforward generalization of the double dark state structure observed in non degenerate four level systems ( see references above ).  Along with Raman dark states the system admits another type of dark states  which have no analogues in  non-degenerate four level  systems.  For certain  polarizations of the rf and probing  fields qualitatively new dark states occur  which do not require  the fulfillment of the two-photon  resonance condition. In the following we will refer to them  as  non Raman dark states to distinguish them from  Raman ones which occur when the two-photon resonance condition takes place.  The origin of non Raman resonances  is  related to the degeneracy of the  configuration:  The rf field and probing field polarization degrees of freedom  form  entangled states in the two dimensional quantum space of the degenerate level, and, as a consequence  interference effects occur in the system  resulting in non Raman resonances. The purpose of the present work is  to study properties of non Raman resonances. We anticipate that non Raman dark states together with  multiple Raman ones can  enlarge the domain of the dark state based physics. We prefer in the theoretical treatments to use  a simple physical  model that can account for the major characteristics of the configuration of Fig. 1(a). For this reason we realize the degenerate four level  system in the atomic hydrogen. We consider  the ground state  relaxation due to  spin-exchange collisions  which is sufficiently good approximation for  a dense hydrogen  gas at room temperatures. Using the full set of density matrix equations we establish properties  of EIT resonances associated with  Raman and non-Raman dark states.

\noindent 

The paper is organized as follows. In the following section   we setup the model and study its  dark states  using the dressed picture formalism.  In Sec. III we establish main properties of Raman and non Raman resonances in a dense hydrogen gas at room temperatures. Our conclusion follows in Sec. IV. 
Density matrix equations for the system are given  in the Appendix. 

\section{Dark states}
 In this work  we  consider  a degenerate four level  atom  with triplet ground levels  $E_1$, $E_2$, and $E_3$ and an excited level $E_4$ interacting with two lasers and one rf field as shown in  Fig. 1.  A twice degenerate level  $E_1$  is spanned by  the pair of orthogonal   vectors $ \mid 1 > $ and $ \mid 1^\prime > $ and the wave functions of the remaining
three levels are $ \mid 2 > $, $ \mid 3 > $ and $ \mid 4 > $.  A laser of frequency    $\omega_c$   is  resonant  with  the    $E_2\rightarrow E_4$   transition and is referred to as the coupling field.  A rf field of frequency   $\omega_r$    is   resonant  with  the    $E_1\rightarrow E_3$   transition   and the absorption spectrum   is obtained by scanning the frequency  $\omega_p$   of the second laser, referred to as the probing field which operates  quasi-resonantly between the  $E_1$ and $E_4$ levels.

\begin{figure}[h]
\includegraphics[width=0.4 \textwidth]{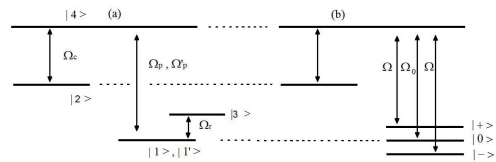} 
\caption{\label{fig:fig1}  (1a) The  degenerate four level system: three metastable and single life-time broadened levels are driven by the coupling ( $\Omega_c$ ),  probing ( $\Omega_p$ ) and  radio frequency  ( $\Omega_r$ )  fields. The lowest level is twice degenerate. (2a) The degenerate four level system in the dressed  picture. Three Raman dark states occur when the coupling and probing fields satisfy the two-photon resonance condition.  Non Raman dark states arise when one of the dressed Rabi frequencies $\Omega$ or $\Omega_0$ is zero. }
\end{figure}

The scanning parameter which we use in the article is the probing field frequency detuning  from the resonance

\begin{equation}
\triangle = \frac{E_4-  E_1}{\hbar}-\omega_p.
\end{equation}
The Rabi frequencies of the coupling  and probing  fields are
denoted by $\Omega_c$ and $\Omega_p$, $\Omega_p^\prime$,
respectively.  The presence of the two Rabi frequencies for the
probing light  is associated with the degeneracy of the level $E_1$.
Not primed and primed Rabi frequencies  are proportional
to  electric  dipole transitions between the ground states  $ \mid 1
> $ and $ \mid 1^\prime > $ and the excited state  $ \mid 4 >$. For
the same reason there are   two  Rabi frequencies for the rf wave  which
we denote by  $\Omega_r$ and  $\Omega_r^\prime$. They are proportional to magnetic dipole transitions within hyperfine levels. The semiclassical Hamiltonian of the quantum system in the rotating wave frame is 
\begin{eqnarray}\label{Ham}
H &=&  \triangle   ( \mid 2 ><2 \mid + \mid 4 ><4\mid )  \label{H}  \\
&-& \frac{1}{2}\left( \Omega_{r} \mid 3 ><1\mid + \Omega^\prime_{r} \mid 3 ><1^\prime \mid + H.c.\right)  \nonumber  \\
  \label{cH}&-& \frac{1}{2}   \left(  \Omega_c \mid 4 ><2\mid +  \Omega_c \mid 2 ><4\mid  \right)   \nonumber \\
  &-& \frac{1}{2} \left(  \Omega_p \mid 4 ><1\mid +   \Omega^\prime_p \mid 4 ><1^\prime \mid + H.c.  \right), \nonumber
\end{eqnarray}
where $H.c.$ stands for the hermitian conjugation.  When treating a strong photon-atom interaction, it is often convenient to introduce the dressed state picture, which allows us to gain more physical insight \cite{Dressed}. In a dressed state picture the  $E_1$ and $E_3$ levels and  the  rf field is  treated as a coupled `''atom + rf field''  system which Hamiltonian  is given  by the expression in the second line of (\ref{H}). Eigenvectors 
\begin{subequations}\label{dspm}
\begin{eqnarray}
\mid - > &=& \frac{1}{2\sqrt{2}} \left( \frac{\overline{\Omega}_r}{\Sigma}  \mid 1 >
+ \frac{\overline{\Omega}^\prime_r }{\Sigma} \mid 1^\prime >+\mid 3 >  \right), \\
\mid 0 > &=&  \frac{\Omega^\prime_r}{2\Sigma}  \mid 1 > -\frac{\Omega_r }{2\Sigma} \mid 1^\prime >, \\
 \mid + > &=& \frac{1}{2\sqrt{2}} \left( \frac{\overline{\Omega}_r}{\Sigma}  \mid 1 > +
\frac{\overline{\Omega}^\prime_r }{\Sigma} \mid 1^\prime >-\mid 3 >\right) 
\end{eqnarray}
\end{subequations}
of the ''atom + rf field''  Hamiltonian form a ladder of triplet as shown in Fig. 1 (b) where the energy difference is defined by the  rf field induced light shift 
\begin{equation}\label{ls}
\Sigma=\frac{1}{2} \sqrt{\mid \Omega_r\mid^2 + \mid \Omega^\prime_r\mid^2}.
\end{equation}
In a similar fashion we may treat the $E_4$ and $E_1$ levels coupled by the the probing  field as  the   ''atom + probing  field'' system which Hamiltonian  is given by the expression in the fourth line of   (\ref{H}). In the dressed state picture this Hamiltonian reads 
\begin{equation}
-\frac{1}{2}\Omega_0 \mid 4 ><0\mid -\frac{ 1}{2}\Omega   \mid 4> \left(  \frac{  < + \mid + < - \mid } {\sqrt{2}}\right) + H.c.,
\end{equation}
where  
\begin{eqnarray}\label{Rabi}
\Omega_0 =  \frac{\Omega_p\Omega_{r}^\prime -\Omega_p^\prime 
\Omega_{r} }{\Sigma }, \ \ \ \Omega = \frac{\Omega_p\overline{\Omega}_{r} +
\Omega_p^\prime\overline{\Omega}^\prime }{\Sigma }.
\end{eqnarray}
are dressed Rabi frequencies. In the absence of relaxations within metastable  states ( the case of ultra cold atomic gas ) dark states of the quantum system may be described by means of the  Schr\"{o}dinger equation. A quantum superposition state   is  decoupled from the coherent interactions if it is orthogonal to  the excited state. Under this condition the  Hamiltonian (\ref{H}) admits three eigenfunctions provided that the frequency detuning is zero or  equal to $\pm \Sigma$. Since  the coupling field is in exact resonance with the transition $\mid 2 >\rightarrow\mid 4>$  the condition $\triangle=0$ is equivalent to the two-photon  resonance condition for  the $\mid 2 > \rightarrow  \mid 0>$ transition. Other two Raman dark states  occur when the  two-photon  resonance conditions for  the $\mid 2 > \rightarrow  \mid \pm>$  transitions are valid. 

\noindent 

We  have  qualitatively new dark states  when one of the Rabi frequencies in  (\ref{Rabi}) is zero. When   $\Omega_0=0$ the state  $\mid 0 >$ is the eigenfunction of the Hamiltonian. This state exists for an arbitrary value  of the frequency detuning. Consequently no two-photon resonance condition is required. In a similar fashion when  $\Omega=0$ we have two non Raman dark states $\mid \pm >$.  To understand   the origin of non Raman dark states  we realize the degenerate four level system in the hydrogen atom  using hyperfine sublevels of the ground states.  The hyperfine level  of the ground state $S_{\frac{1}{2}}$ with the total angular momentum $F$ is  spanned by wave functions $\mid S, F, m >$ where the magnetic quantum number $m$   enumerates  Zeeman sub-levels of the F-hyperfine level. We make the following choice

\begin{subequations}\label{Zeeman}
\begin{eqnarray}
    \mid 1> &=& \mid S,1;  1 >, \\
    \mid 1^\prime > &=& \mid S, 1; -1 >, \\
    \mid 2>  &=& \mid S, 1; 0 >, \\
    \mid 3>  &=& \mid S, 0; 0 >. 
 \end{eqnarray}
\end{subequations}
As the excited state  we choose the Zeeman sublevel of the hyperfine level with the total angular momentum $F=0$ in the $P_{\frac{1}{2}}$ fine structure
\begin{eqnarray}
  \mid 4>  &=& \mid P, 0; 0 >.
\end{eqnarray}
Transitions from  the ground state into  the excited state are described by the electric dipole Hamiltonian $-\vec{d}\vec{\cal{E}}(t)$, where $\vec{d}$ is the electric dipole  operator of the atom and
\begin{equation}
  \vec{\cal{E}}(t) = \frac{1}{2} \vec{\mathcal{E}} e^{-i\omega_p t }+\frac{1}{2}\vec{\mathcal{E}}^* e^{i\omega_p t }
\end{equation}
is the electric field. The  Rabi frequencies 
\begin{subequations}\label{electrical}
\begin{eqnarray}
  \Omega_{p}  &=& \frac{1}{\hbar}  < P, 0,0\mid  \vec{d} \vec{\cal{E}}  \mid  S, 1;  1 >, \\
 \Omega^\prime_{p}  &=& \frac{1}{\hbar}  < P, 0,0\mid  \vec{d} \vec{\cal{E}}  \mid  S, 1; - 1 >
\end{eqnarray}
\end{subequations}
take the form 
\begin{eqnarray}\label{pr}
  \Omega_{p}  = \frac{d}{\hbar} \frac{ \mathcal{E}_x+ i  \mathcal{E}_y}{\sqrt{2}},  \ \ \ \   \Omega^\prime_{p}  = \frac{d}{\hbar} \frac {\mathcal{E}_x- i  \mathcal{E}_y}{\sqrt{2}}.
\end{eqnarray}
$\mathcal{E}_x$ and $\mathcal{E}_y$ are $x$ and $y$ components of the electric field and $d$ is the atomic  dipole moment. Levels within the same hyperfine structure have the same parity  and the rf-field couples $E_1$ and $E_3$ levels by the magnetic dipole Hamiltonian $-\vec{m}\vec{\cal{H}}(t)$, where $\vec{m}$ is the magnetic  dipole  operator of the atom and $\vec{\cal{H}}(t)$ is the magnetic component of the rf field. Repeating for the rf field  the above derivations we obtain 
\begin{eqnarray}\label{rf}
  \Omega_{r}  = \frac{m}{\hbar} \frac{ \mathcal{H}_x+ i  \mathcal{H}_y}{\sqrt{2}}, \ \ \ \ 
 \Omega^\prime_{r} = \frac{m}{\hbar} \frac {\mathcal{H}_x- i  \mathcal{H}_y}{\sqrt{2}}.
\end{eqnarray}
Here  $m$ is the magnetic moment which is proportional to the  Bohr magneton. Inserting (\ref{pr}) and (\ref{rf}) in (\ref{Rabi})  we finally obtain  
\begin{eqnarray}\label{pr1}
  \Omega_0 = \frac{i d}{\hbar}   \frac{ \mathcal{E}_y \mathcal{H}_x - \mathcal{E}_x \mathcal{H}_y }{ \mid \vec{\mathcal{H}} \mid }, \ \ \ 
  \Omega   = \frac{d}{\hbar}   \frac{ \mathcal{E}_x \overline{\mathcal{H}}_x+ \mathcal{E}_y \overline{\mathcal{H}}_y }{ \mid \vec{\mathcal{H}}\mid }.    
\end{eqnarray}
The dressed Rabi frequencies are defined by the scalar product and by the $z$ component of the vector product between the polarization vectors of the rf and probing fields. In the case of  linear polarizations   non Raman  dark states occur when the probing field polarization  is parallel or perpendicular to that of the  rf field.

\section{Propagation of weak probe field }
  To quantify the properties of EIT resonances associated with Raman and non Raman dark states for an atomic gas at room  temperatures  we examine the linear response of the system ( Fig.1 a ) using the full set of density matrix equations given in the Appendix A. In a sufficiently  dense hydrogen gas ground state relaxations are limited  by  spin-exchange  mechanism \cite{Bender, spin1}.  The evolution of the atom  due to  the spin-exchange scattering is described by equations (\ref{ex}). We will use linear approximation for the spin-exchange relaxation which is valid  when the average spin of the gas is close to zero \cite{spin2}.  The spin- exchange rate $\gamma$ is a function of the  density and  the temperature of a gas. In this paper a ratio of $\gamma/\Gamma=10^{-4}$ is used.  We assume that  the rf field is linear polarized along the $x$ direction ($\mathcal{H}_y=0$ ) and the probing electric field is linear polarized in the $x0y$ plane 
\begin{eqnarray}\label{angle}
   \mathcal{E}_x   =    \mathcal{E}\cos\psi, \ \ \    \mathcal{E}_y   =   \mathcal{E} \sin\psi, 
\end{eqnarray}
where $\psi$ is the angle between $ \vec{\mathcal{E}}$ and  $ \vec{\mathcal{H}}$. 
\\
\\
 The linear response $\sigma$ of atoms  to a weak probing  field  defines the dielectric polarization of the medium
\begin{eqnarray}\label{polar}
  \mathcal{P}_x  =   2 d\varrho  ( \sigma_{41^\prime}+\sigma_{41} ),  \ \   \mathcal{P}_y  =   2i d\varrho ( \sigma_{41^\prime}-\sigma_{41} ),
\end{eqnarray}
where $\varrho$ is the number of atoms per unit volume in the interaction region. Using solutions of the Liouville equation  (\ref{Liu})  we find  that   non diagonal components of the dielectric  susceptibility tensor are zero. As a result we have  
\begin{eqnarray}\label{polar1}
\mathcal{P}_x  = \chi_{x} \mathcal{E}_x, \ \ \ \ 
\mathcal{P}_y  = \chi_{y} \mathcal{E}_y.
\end{eqnarray}
When the rf field is switched off we have  $\chi_x=\chi_y$ and the medium becomes isotropic for the probe field. The $E_3$ level is decoupled from other  ones  and we arrive at the  Lambda system with single EIT resonance which appears at the zero frequency detuning  as shown in the Fig. 2(a).

\begin{figure}[h]
\includegraphics[width=0.4 \textwidth]{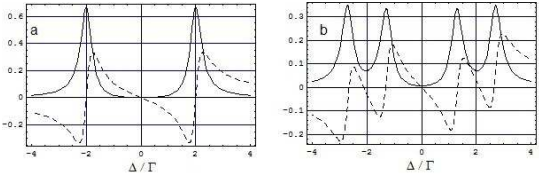} 
\caption{\label{fig:fig2}  Imaginary ( solid lines ) and real ( dashed lines ) parts of the  susceptibility function $\chi(\psi )$ in the units of $\lambda$ for the degenerate four level system. (a) the case when the rf field is switched off. The same is true for nonzero rf field when the polarization vector  of the rf field is parallel to that of the probing field. (b) Triple dark state structure appears when   the rf field  and probing  field polarizations are perpendicular to each other. Parameters are $\Omega_c=4\Gamma$, $\Omega_r=\Gamma$.}
\end{figure}

\noindent

To study propagation of a weak probing field  in the presence of the rf field  we  first  define absorption and refraction functions in an anisotropic  medium. The energy dissipation rate in the probing field per unit volume in the atomic medium is defined by the  divergence of the energy flux density \cite{Landau} and is given by 
\begin{equation}
Q = \frac{ \mathcal{E}^2  }{8\pi T } F 
\end{equation}
where the dimensionless function 
\begin{equation}\label{diss}
F= 8\pi^2 \Im\left(  \chi_{x}\cos^2\psi  + \chi_{y} \sin^2\psi \right) 
\end{equation}
determines the ratio of the absorbed  energy  during the period  of the oscillation $T=\frac{2\pi}{\omega_p}$ to the free-space energy density of the probing field. In the regions near EIT resonances where the medium can be considered as transparent we may define the group velocity.  We neglect the imaginary part of dielectric susceptibilities and define real wave vector $\vec{k}$ which determines the propagation direction. Since magnetic transitions between ground states and the excited state are forbidden the probing magnetic  field coincides with the  magnetic induction. In this non magnetic case the Maxwell equations are reduced to the Fresnel's ones 
\begin{equation}\label{fren}
(k^2-\frac{\omega_p^2}{c^2}) \vec{\mathcal{E}}- (\vec{k}, \vec{\mathcal{E}})\vec{k}=4\pi \frac{\omega_p^2}{c^2}\vec{\mathcal{P}}.
\end{equation}
 Assuming that the probing  light propagates in xOy plane  
\begin{equation}
k_x=k\sin\phi, \ \ \ \ k_y=k\cos\phi
\end{equation}
 from (\ref{fren}) we obtain  dispersion relations 
\begin{eqnarray}\label{refr0}
k^2  = \frac{\omega_p^2}{c^2}
  \frac{ \epsilon_x^2\cos^2\psi +\epsilon_y^2\sin^2\psi }{ \epsilon_x\cos^2\psi +\epsilon_y\sin^2\psi}
\end{eqnarray}
and 
\begin{equation}
\cot\phi\tan\psi =-\frac{ \epsilon_x}{ \epsilon_y}.
\end{equation}
Here $\epsilon_a=1+4\pi \chi_a$ is the electric permittitivy. When the dielectric susceptibilities are much smaller than unity we may rewrite (\ref{refr0}) in the form 
\begin{eqnarray}\label{refr}
k  = \frac{\omega_p}{c} ( 1 +2\pi \Re\left(  \chi_{x}\cos^2\psi  + \chi_{y} \sin^2\psi \right) ).
\end{eqnarray}
Using (\ref{diss}) and (\ref{refr}) we conclude  that absorption and refraction rates are proportional to the imaginary and the real parts of the susceptibility function 
\begin{eqnarray}\label{sus}
\chi (\psi ) =  \chi_{x}\cos^2\psi  + \chi_{y} \sin^2\psi .
\end{eqnarray}
An atomic electron subjected to  the $x$ -polarized  rf field   will oscillate in the plane which is transverse to the $x$ axis. Consequenly  no atomic dipole moment in $x$ direction can be induced by the rf field. As a result the x component of the susceptibility  $\chi_x$ will be  independent of  the rf field in the first order perturbation theory. From (\ref{sus}) we conclude that when the rf and probing fields are  polarized in the same direction ( $\psi=0$ ) the degenerate four level system admits single EIT resonance similar to that of the Lambda system  ( see Fig 2 (a) ).  On the other hand when the polarization vectors of the fields are  perpendicular to each other  the quantum system acquires the triple dark state structure as shown in Fig. 2(b).  The shape and position of EIT resonances can be manipulated by adjusting the angle $\psi$.

\begin{figure}[h]
\includegraphics[width=0.4 \textwidth]{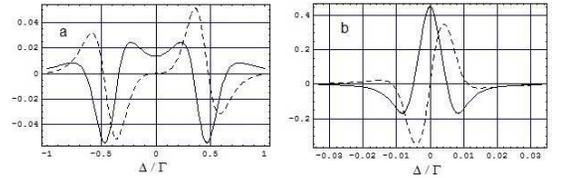} 
\caption{\label{fig:fig3}  Imaginary ( solid lines ) and real ( dashed lines ) parts of the  susceptibility function $\delta \chi$ in the units of $\lambda$ for the  degenerate four level system. (a) Triple dark state  structure occurs  at  $\Omega_c=\Gamma$ and $\Omega_r=0,1 \Gamma$;  (b) Double  dark state structure occurs  at  $\Omega_c=0.1\Gamma$ and $\Omega_r=0,01 \Gamma$}
\end{figure}

Since the maximum power of the rf ( or  microwave ) field is limited due to experimental difficulties it is of particular interest to examine the influence of a weak rf field on optical  properties of a medium.  To analyse EIT resonances in this case  it is instructive to represent the susceptibility function (\ref{sus}) in the following form
\begin{eqnarray}\label{decomp}
\chi (\psi ) = \chi ( 0 )+  \delta \chi \sin^2\psi  
\end{eqnarray}
where $\chi ( 0 )$ describes absoption and refraction effects in the absence of the rf field ( which is the Lambda system, as we have  mentioned above ) and   $\delta \chi$ includes effects induced by the rf field. The contribution of the rf field to  absorption and refraction profiles are shown in Fig.3.  Depending on the amplitude of the fields, these resonances can have  either a double or a triple dark state structure. Fig.3  depicts an   important feature of these resonances: The refraction index acquires large values in regions where the absorption function is close to zero or even  negative  ( the region of the optical gain ). Such property of the rf field induced resonances can be used to improve  the refraction index  of the medium. 

\noindent

Imposing the transparency condition $\Im \left( \chi (\psi))\right) =0$ on  (\ref{decomp}) we obtain the polarization  geometry  which leads to the emergence of non Raman resonances 
\begin{eqnarray}\label{geom}
 \sin^2\psi  = -\frac{\chi ( 0 )}{\delta \chi}.  
\end{eqnarray}
When  the probe field is in the exact resonance with the $E_1\rightarrow E_4$ transition ( that is  $\triangle=0$ ) we have 
\begin{eqnarray}\label{I1}
\Im \left(\chi (\psi ) \right) & =&   \frac{4\lambda \gamma \Gamma^2}
{8\gamma \Gamma^2+3\Omega_c^2\Gamma+10 \Omega_c^2 \gamma } \nonumber \\
&\times & \left(  \frac{\Gamma \Omega_c^2-2\gamma \Omega_c^2+4 \Gamma^2\gamma}{\gamma ( \Omega_c^2+2 \Gamma\gamma)} \frac{\Omega_r^2 }{\Omega_c^2}\sin^2\psi -1 \right) .
\end{eqnarray}
Here  $\lambda= \frac{\sqrt{2}\varrho d^2}{\hbar \Gamma}$ and we have preserved terms up to the second order in $\frac{\Omega_r}{\Omega_c}$. At $\psi=0$ the imaginary part of the susceptibility function (\ref{sus}) is negative and we have the optical gain for the probing field. The energy of the coupling light is transformed into the one  of the probing  light by means of the spin-exchange mechanism. The physics of this phenomenon  is similar to that of the  Hanle effect where the polarization of an  incident laser light is changed by a  constant magnetic field \cite{hanle}. In our case the probing field polarization which is directed  along the $z$ -axis  is transformed into the linear polarization in the $xOy$ plane. By the virtue of  (\ref{geom}) and (\ref{I1}) we obtain  the  polarization configuration 
\begin{eqnarray}\label{NR}
\sin^2\psi= \frac{\gamma( \Omega_c^2+2 \Gamma\gamma)}{\Gamma \Omega_c^2-2\gamma \Omega_c^2+4 \Gamma^2\gamma}
 \frac{\Omega_c^2 }{\Omega_r^2}
\end{eqnarray}
at which  non Raman resonances occur. From (\ref{I1}) we observe that  non Raman  resonances are critical points at which  the probing field absorption turns into the optical gain.  

\noindent

Although an rf ( or microwave ) source is more readily available and easier  to control in comparison with an extra laser field  there are practical limitations on the  power of the rf field. This creates  difficulties in the experimental observations  of rf field induced light shifts  in Lambda type systems. To observe narrow double dark state resonance ( which is the EIT resonance associated with Raman dark states in non-degenerate four level system ) one should overcome the Doppler broadening that attenuates the narrow features of optical spectra. Using special Doppler free geometry a narrow double dark state was observed in Rb atomic vapor in  \cite{DDS2}. Non Raman resonances are not defined by an rf field induced light shift, they have geometrical origin rather than ''energetic''.  We anticipate that non Raman resonances may provide new possibilities  for experimental observations of  rf  filed ( or microwave ) effects in multilevel systems.  In particular non Raman resonances may be of interest in the metrology science for  measuring of an rf field amplitude with high accuracy \cite{Must}. The presence of dipole transitions in the optical spectrum  makes  alkali atoms more suitable for an experimental  implementation of the configuration proposed in this work.  This can be achieved for example by realization of the degenerate four level  system  within the Rb $D_1$ absorption  line. The $\mid 1>$, $\mid 1^\prime>$ and $\mid 2>$  states can be set up with three $F=1$ Zeeman sub-levels similarly with the H atom ( see the equation (\ref{Zeeman}) ), and  the remaining $\mid 3>$ state is set up with the $F=2$, $m=0$  Zeeman state. We expect that general properties of EIT resonances for atoms with single outer $s$ electron will resemble the ones which we have established  for the atomic hydrogen.   

\section{Conclusions}
This article provides a detailed theoretical treatment of EIT resonances in the configuration of Fig.1(a).  The degeneracy in the ground state  gives rise to the appearance of new type of dark states in the system.  Along with two-photon Raman resonances ( the triple dark state structure ) there are  non Raman ones which have geometrical origin.  We realize the configuration in the atomic hydrogen and show that non Raman dark states are defined by the geometry of the rf and probing field's   polarizations.   Using the dressed picture formalism we show that for ultra-cold atoms non Raman resonances are determined by  scalar and  vector  products of the polarization vectors. Properties of Raman and non Raman  resonances for a dense hydrogen gas at room temperatures   are studied  by examing the linear response of atoms  to the weak probe field. We take into account  the spin-exchage relaxation  in the ground state of the atomic hydrogen.   Using solutions of the density matrix equations we find that two components of the diagonal susceptibility tensor differ by a factor which is proportional to the magnitude of the rf field. As a result a hydrogen gas becomes anisoptropic for the probing field in the presence of the rf field . Absorption and refraction properties of the medium  is defined by the imaginary and the real parts of the susceptibility function (\ref{sus}). The Lambda type EIT resonance  appears in the case when the probing field is polarized in the same direction as the rf one.  On the other hand  when the  probing field polarization is perpendicular to that of the rf field  the triple dark state structure arises  which is the extension  of the double dark structure observed in non degenerate four level systems  \cite{DDS1,Tripod}. It is possible to represent  optical properties of the medium as the sum of the Lambda type EIT resonance and an rf field induced resonace. The latter is proportional to the square of the sine function which argument represents the angle between the probing and the rf field's polarizations ( see Eq. (\ref{decomp})  ). We observe an interesting feature of the rf field induced resonance: The refraction index  reaches maximal values in regions where the   absorption rate is close to zero or even negative ( the region of the optical gain ). When the probing field is resonant  we give the analytic expression for EIT resonances and establish the connection of non Raman dark states  with the polarization geometry. 

\begin{acknowledgments}
 We are grateful to Dr. Mustafa Cetintas who  pointed out the importance of new   microwave amplitude  measurement methods  in the metrology science. 
\end{acknowledgments}

\appendix 
\section{Linear response}
The linear response $\sigma$  of the atomic system  to the weak probe field obeys    the  Liouville  equation
\begin{equation}\label{Liu}
\frac{ d^{ex} \sigma
}{dt}+ \frac{d^{sp} \sigma}{dt}= i [H_0,\sigma] +  i [V,\rho].
\end{equation}
Here 
\begin{eqnarray}
V &=& -\frac{1}{2} ( \Omega_p   \mid 4 ><1\mid + \Omega^\prime_p  \mid 4 ><1^\prime \mid + H.c )
\end{eqnarray}
is the  the probe field potential where the  Rabi frequencies due to (\ref{electrical}) and  (\ref{angle})  take the form
\begin{eqnarray}
\Omega_p  = \frac{d}{\sqrt{2}\hbar } \mathcal{E} e^{i\psi}, \ \ \ \Omega^\prime_p  = \frac{d}{\sqrt{2}\hbar } \mathcal{E} e^{-i\psi}.
\end{eqnarray}
Dropping the  potential $V$ from  (\ref{Ham}) we arrive at  the Hamiltonian $H_0$  of ``atom + coupling field + rf field'' system.  The evolution of the atom due to the spontaneous light  emission is given by
\begin{subequations}
\begin{eqnarray}
  \frac{d^{sp} \sigma_{44}}{dt}  &=& -\Gamma \sigma_{44} \\
 \frac{d^{sp} \sigma_{\beta\beta}}{dt}
    &=& \frac{1}{4} \Gamma  \sigma_{44} \\
 \frac{d^{sp} \sigma_{4\beta}}{dt }&=& -  \frac{1}{2}\Gamma \sigma_{4\beta},
\end{eqnarray}  
\end{subequations}
where  $\beta=1,1^\prime,2,3$ and $\Gamma$ is the decay rate of the upper  state. Here for simplicity we assume that the ratios of the population decay from the excited state to different ground states are equal. We have used the symbol $\frac{d^{sp}}{dt}$ to distinguish the evolution of $\sigma$ due to spontaneous emissions ( $sp$ ) from the evolution of the density matrix due to spin-exchange collisions  ( $ex$ ) which in the  linear approximation reads 
\begin{subequations}\label{ex}
 \begin{eqnarray}
   \frac{d^{ex} \sigma_{11}}{dt}  &=& -   \frac{\gamma}{2} (\sigma_{11}+\sigma_{1^\prime 1^\prime} -\sigma_{22}-\sigma_{33}) \\
   \frac{d^{ex} \sigma_{22}}{dt}  &=& - \frac{\gamma}{2} (3\sigma_{22}-\sigma_{1^\prime 1^\prime} -\sigma_{11}-\sigma_{33}) \\
    \frac{d^{ex} \sigma_{1^\prime 1^\prime}}{dt}  &=& -   \frac{\gamma}{2} (\sigma_{11}+\sigma_{1^\prime 1^\prime} -\sigma_{22}-\sigma_{33}) \\
        \frac{d^{ex}\sigma_{33}}{dt}  &=& - \frac{\gamma}{2} (3\sigma_{33}-\sigma_{1^\prime 1^\prime} -\sigma_{11}-\sigma_{22})\\
       \frac{d^{ex}\sigma_{21}}{dt}   &=&-\gamma (\sigma_{21}-\sigma_{1^\prime 2} ) \\
        \frac{d^{ex} \sigma_{1^\prime 2}}{dt}  &=&-\gamma (\sigma_{1^\prime 2} -
       \sigma_{21}) \\
        \frac{d^{ex} \sigma_{1^\prime 1}}{dt} &=&-2 \gamma \sigma_{1^\prime 1}\\
         \frac{d^{ex} \sigma_{3\alpha}}{dt} &=&-\gamma  \sigma_{3\alpha}.
 \end{eqnarray}
 \end{subequations}
Here $\alpha=1,1^\prime,2$ and $\gamma$ is the spin exchange rate. In the general case  evolution equations due to spin-exchange collisions contain also  quadratic terms in the density matrix components which are  proportional to the total spin of an atomic gas \cite{Bender,spin1}. 

\noindent 

The steady state  of an atomic gas  prepared by  the  coupling field and the rf field   satisfy  the equation
\begin{equation}
\frac{ d^{ex} \rho}{dt}+ \frac{d^{sp} \rho}{dt}= i [H_0,\rho].
\end{equation}
and is given by the density matrix whose non zero matrix elements are   
\begin{subequations}
\begin{eqnarray}
\rho_{11}  &=& \frac{2 \gamma (\Gamma^2+\Omega_c^2)+ \Omega^2_c\Gamma}{8\Gamma^2\gamma+10\Omega_c^2\gamma +3\Gamma\Omega_c^2  } \\
\rho_{22}  &=& \frac{2 \gamma (\Gamma^2+\Omega_c^2) }{8\Gamma^2\gamma+10\Omega_c^2\gamma +3\Gamma\Omega_c^2  } \\
\rho_{44} &=& \frac{2 \gamma \Omega_c^2}{8\Gamma^2\gamma+10\Omega_c^2\gamma +3\Gamma\Omega_c^2  } \\
 \rho_{42}&=& i \frac{\Gamma}{\Omega_c}\rho_{44}, \ \ \ \rho_{24}= -i \frac{\Gamma}{\Omega_c}\rho_{44}
\end{eqnarray}
\end{subequations}
and $ \rho_{11} = \rho_{1^\prime 1^\prime }=\rho_{33}$.

\end{document}